\documentclass[manuscript]{aastex63}

\received{June 1, 2019}
\revised{January 10, 2019}
\accepted{\today}
\submitjournal{ApJ}

\shorttitle{Confirming the Explosive Outflow in G5.89 with ALMA}
\shortauthors{Zapata et al.}
\usepackage{graphicx}

\newcommand{\dechms}[4]{$#1^{\rm h}#2^{\rm m}#3\mbox{$^{\rm s}\mskip-7.6mu.\,$}#4$}

\newcommand{\decdms}[4]{$-#1^{\circ}#2'#3\mbox{$''\mskip-7.6mu.\,$}#4$}

\begin{document}

\title{Confirming the Explosive Outflow in G5.89 with ALMA}

\correspondingauthor{Luis A. Zapata}

\email{l.zapata@irya.unam.mx}

\author{Luis A. Zapata}
\affil{Instituto de Radioastronom\'\i a y Astrof\'\i sica, Universidad Nacional Aut\'onoma de M\'exico, P.O. Box 3-72, 58090, Morelia, Michoac\'an, M\'exico}

\author{Paul T. P. Ho}
\affil{Academia Sinica Institute of Astronomy and Astrophysics, PO Box 23-141, Taipei, 10617, Taiwan}
\affil{East Asian Observatory, 666 N. A'ohoku Place, Hilo, Hawaii 96720, USA}

\author{Manuel Fern\'andez-L\'opez}
\author{Estrella Guzm\'an Ccolque}
\affiliation{Instituto Argentino de Radioastronom\'\i a (CCT-La Plata, CONICET; CICPBA), C.C. No. 5, 1894, Villa Elisa, Buenos Aires, Argentina}

\author{Luis F. Rodr\'\i guez}
\affil{Instituto de Radioastronom\'\i a y Astrof\'\i sica, Universidad Nacional Aut\'onoma de M\'exico, P.O. Box 3-72, 58090, Morelia, Michoac\'an, M\'exico}
\affil{Mesoamerican Centre for Theoretical Physics, Universidad Aut\'onoma de Chiapas, Carretera Emiliano Zapata Km. 4 Real del Bosque, 29050 Tuxtla Guti\'errez, Chiapas, M\'exico}

\author{Jos\'e Reyes-Vald\'es}
\affil{Centro de Investigación en Matem\'aticas Aplicadas, Universidad Aut\'onoma de Coahuila, Camporredondo S/N, Saltillo, Coahuila, CP 25115.}

\author{John Bally}
\affil{Astrophysical and Planetary Sciences Department University of Colorado, UCB 389 Boulder, Colorado 80309, USA}

\author{Aina Palau}
\affil{Instituto de Radioastronom\'\i a y Astrof\'\i sica, Universidad Nacional Aut\'onoma de M\'exico, P.O. Box 3-72, 58090, Morelia, Michoac\'an, M\'exico}

\author{Masao Saito}
\author{Patricio Sanhueza}
\affil{National Astronomical Observatory of Japan, National Institutes of Natural Sciences, 2-21-1 Osawa, Mitaka, Tokyo 181-8588, Japan}
\affil{ Department of Astronomical Science, SOKENDAI (The Graduate University for Advanced Studies), 2-21-1 Osawa, Mitaka, Tokyo 181-8588, Japan}

\author{P. R. Rivera-Ortiz}
\affil{Univ. Grenoble Alpes, CNRS, Institut de Planétologie et d’Astrophysique de Grenoble (IPAG), 38000 Grenoble, France}

\author{A. Rodr\'\i guez-Gonz\'alez}
\affil{Instituto de Ciencias Nucleares, Universidad Nacional Aut\'onoma de M\'exico, Ap. 70-543, 04510 D.F., M\'exico}




\begin{abstract}
The explosive molecular outflow detected decades ago in the Orion BN/KL region of massive star formation was considered to be a bizarre event. This belief was strengthened 
by the non detection of similar cases over the years with the only exception of the marginal case of DR21. 
Here, we confim a similar explosive outflow associated with the UCH$_{\rm II}$ region G5.89$-$0.39 that indicates that
this phenomenon is not unique to Orion or DR21.  Sensitive and high angular resolution ($\sim$ 0.1$''$) ALMA 
CO(2$-$1) and SiO(5$-$4) observations show that the molecular outflow in the massive star forming region G5.89$-$0.39 is indeed an explosive outflow 
with an age of about 1000 yrs and a liberated kinetic energy of 10$^{46-49}$ erg. Our new CO(2$-$1) ALMA observations revealed over 30 molecular filaments, with Hubble-like expansion motions,  
pointing to the center of UCH$_{\rm II}$ region. In addition, the SiO(5$-$4) observations reveal warmer and strong shocks very close to the origin of the explosion, 
confirming the true nature of the flow. A simple estimation for the occurrence of these explosive events during the formation of the massive stars 
indicates an event rate of once every  $\sim$100 yrs, which is close to the supernovae rate.  
\end{abstract}



\section{Introduction} \label{sec:intro}

The explosive outflows (with a kinetic energy injected that reaches the 10$^{47-49}$ erg range) are suggested to be powered by the liberation of gravitational energy associated 
with the formation of a close-by stellar massive binary or maybe a protostellar merger \citep{ball2005,zap2017,ball2017}.  The explosive molecular outflows are composed of 
tens of narrow straight molecular filament-like ejections with clear Hubble-like velocity increments that point back approximately to a common origin, and with a nearly 
isotropic configuration \citep{zap2009, ball2017,ball2020}. \citet{riv2019} showed that the duration of such explosive outflows should be around some thousand years. 
As more cases like Orion-KL or DR 21 come to light we would know in what extent the dynamical interactions are a key ingredient in oder to form massive stars. Sensitive searches 
using the recently finished ALMA observatory will be very important to discover more explosive outflows and runaway stars in close-by massive star forming regions. 

G5.89$-$0.39 or  W28A2 is a massive star forming region located at a distance of 2.99$^{+0.19}_{-0.17}$ kpc \citep{sat2014}. This region contains a bright expanding shell-like UCH$_{\rm II}$ region
with a dynamical age of 600$^{+250}_{-125}$ years, estimated from its expansion rate \citep{aco1998}. \citet{fel2003} proposed that a massive young O5 V star, revealed by 
their near infrared NACO-VLT observations, is energizing the UCH$_{\rm II}$ region. This object, however, is offset by about 1$''$ to the northwest from the center of the shell-like UCH$_{\rm II}$ region.
\citet{fel2003} also proposed that the young O5 V star may have migrated from the center of the UCH$_{\rm II}$ region to the present place with a velocity of about 10 km s$^{-1}$ some 1000 yrs ago.
A second infrared object is also located at the edge of the UCH$_{\rm II}$ region that is called Puga's star \citep{pug2006} and is related with a northeast optical outflow.  
The nature of the molecular outflow in this region is not clear \citep{zap2019}. \citet{har1988} reported that the molecular outflow located in G5.89$-$0.39 is one of the most powerful outflows in 
our galaxy with a kinetic energy of $\sim$10$^{49}$ erg. However, more recent estimations place the energetics of this outflow in a range between 10$^{46-48}$ erg \citep{aco1998,kla2006}.
Using Chandra observations, \citet{ham2016} reported the presence of a gamma-ray source (HESSJ1800-240B) associated with G5.89$-$0.39 that maybe is related with the outflow arising from here.
Very recently, \citet{zap2019} using Submillimeter Array (SMA) observations, detected the presence of six explosive 
filaments pointing directly to the center of the UCH$_{\rm II}$ region. They deduced the presence of an explosive outflow and suggested a relation between the 
UCH$_{\rm II}$ region and the flow, but their study was not conclusive.   

\begin{figure*}
\center
\includegraphics[scale=0.64, angle=0]{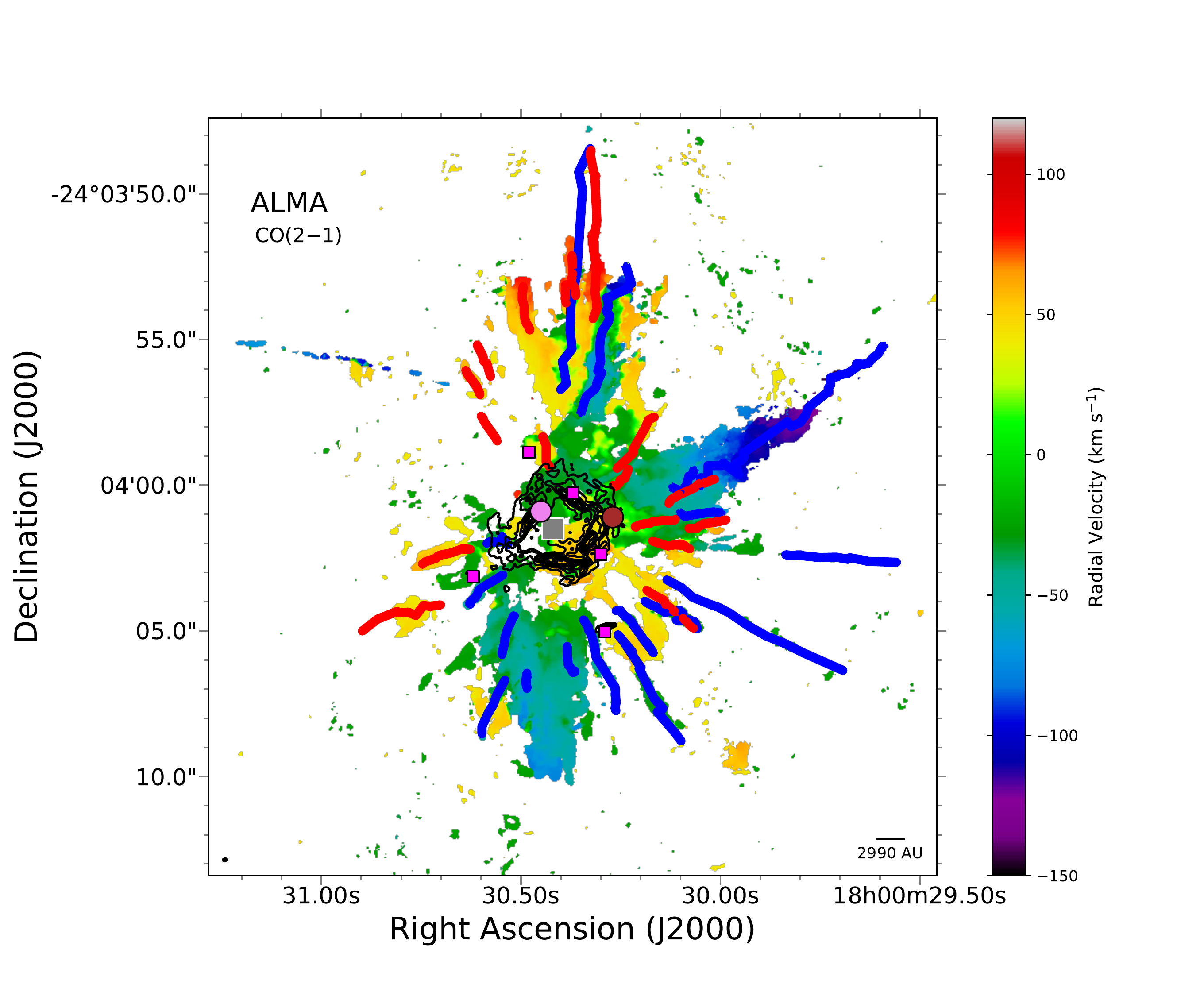}
\caption{ \scriptsize ALMA CO(2$-$1) moment one map overlaid with the approaching (blue) and receding (red) explosive filaments in the G5.89$-$0.39 outflow and the 1.3 mm continuum emission 
in contours that are tracing the UCH$_{\rm II}$ region in the central part.  At these wavelengths emission from UCH$_{\rm II}$ region is still dominated by free-free emission \citep{hun2008}.
To compute this map, we integrated in radial velocities from $-$150 to  $-$20 km s$^{-1}$ for the blueshifted emission, 
while from $+$40 to $+$120 km s$^{-1}$ for the redshifted emission. The contours range from 10\% to 70\% of the peak emission, in steps of 10\%. The peak of the millimeter continuum emission is 34 mJy Beam$^{-1}$.  
The position and velocity of every condensation can be seen in the 3D image presented in Figure 3.   
The half-power contour of the synthesized beam of the line image is shown in the bottom-left corner. The LSR radial velocity scale-bar (in km s$^{-1}$) is shown at the right. In the bottom 
right-corner is also shown the spatial scale at a distance of 2.99 kpc.   The location of the sources named Feldt's star \citep{fel2003} (pink circle) and Puga's star \citep{pug2006} (brown circle) are shown  
at the center of the explosive outflow. The magenta squares mark the location of the objects reported with the SMA \citep{hun2008}. The grey square marks the origin of the outflow.
\label{fig:f1}}
\end{figure*}    

Here, using much more sensitive\footnote{A factor of 40 better in the rms noises, compared with the previous SMA observations, \citet{zap2019}}~spectral line ALMA (Atacama Large Millimeter/Submillimeter Array) 
CO(2$-$1) and SiO(5$-$4) observations together with a much better beam area\footnote{Almost a factor of 70, again compared with the previous SMA observations} 
provided us a better fidelity in the spectral and continuum images to study in more detail the intriguing molecular outflow in G5.89$-$0.39.

 \section{Observations} \label{sec:obs}

The ALMA observations of G5.89$-$0.39 were carried out in 2019 September 5th as part of the Cycle 6 program 2018.1.00513.S (PI: Paul Ho). These observations were split in two
sessions, one with 49 antennas and the other with 45 antennas, both using antennas with diameters of 12 m, yielding baselines with projected lengths from 38 to 3600 m (29–2769 k$\lambda$).
As the primary beam at this frequency (Band 6) has a Full Width Half Maximum (FWHM) of about 25$''$, the outflow from G5.89$-$0.39 was covered with a single pointing at 
the-sky-position $\alpha_{J2000.0}$ = \dechms{18}{00}{30}{388}, and $\delta_{J2000.0}$ = \decdms{24}{04}{00}{20}. The total integration time on source that includes the two sessions is 78 min. 
The maximum recoverable scale for these observations is 2.5$''$.

\begin{figure}
\centering
\includegraphics[scale=0.63, angle=0]{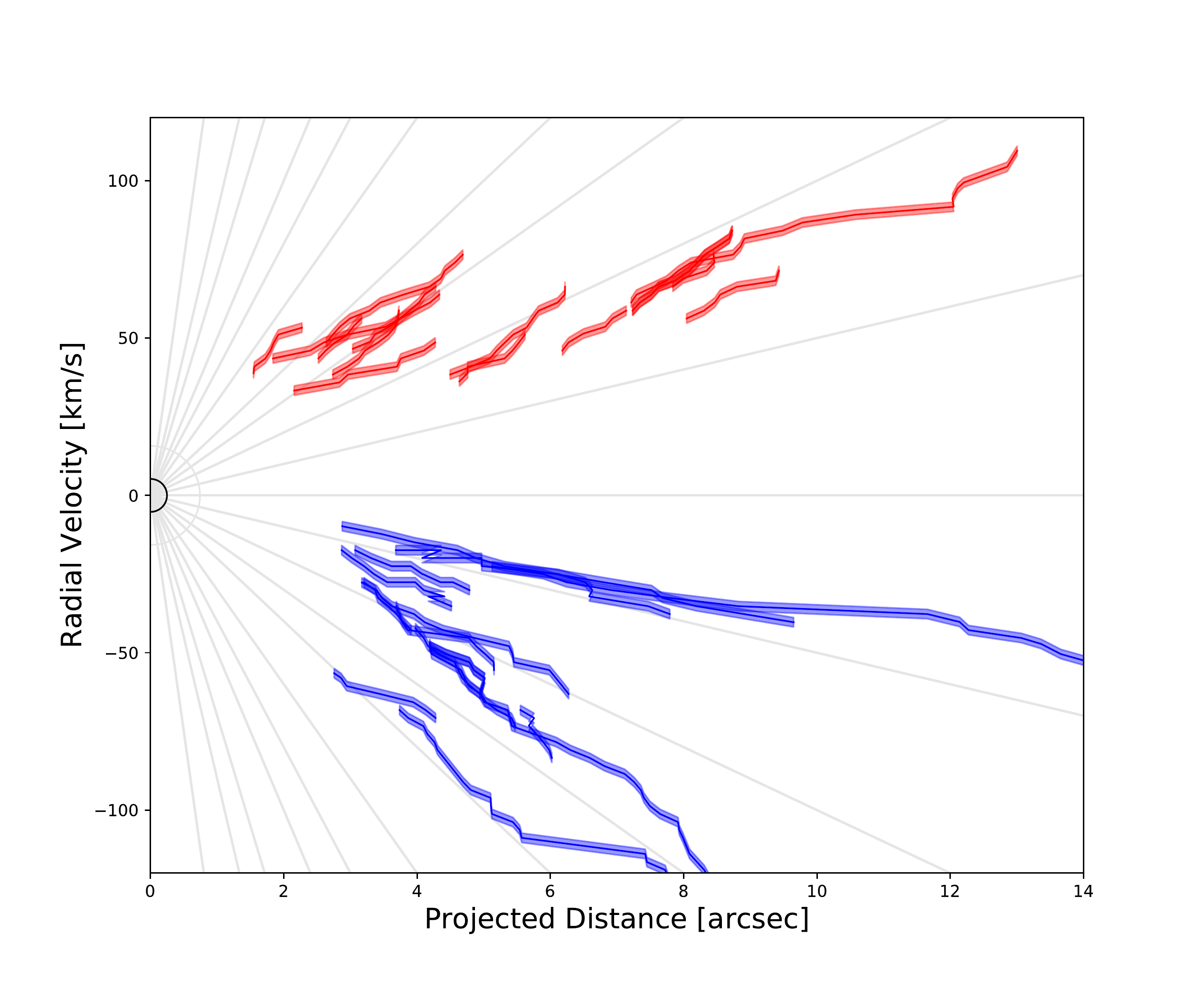}
\caption{ \scriptsize Position-Velocity diagram of the explosive CO(2$-$1) filaments in the G5.89$-$0.39 outflow found with the present ALMA observations.  
Here, we have only included the maximum velocities in a range between $\pm$120 km s$^{-1}$, there are some molecular filaments that extend outside of this window
as can be noted in Figure 1. We have colored the approaching filaments in blue while the receding ones in red. In each filament the shaded blue and 
red color represent the error area that is about 0.1$''$ for the distance on-the-sky, and 1.2 km s$^{-1}$ for the radial velocity. As in the Orion BN-KL 
and DR21 explosive outflows, all velocities vary linearly with distance with no clear deceleration. Almost all filaments seem to converge to a systemic 
velocity range $+$12 to $-$10 km s$^{-1}$. The cloud velocity of most of the detected molecular lines is around 9 km s$^{-1}$ \citep{hun2008,su2009}.  All grey lines begin at R=0 
and a radial velocity of 0 km s$^{-1}$.
\label{fig:f2}}
\end{figure}

The weather conditions were excellent during these observations, with an average system temperature around 90 K and with an average precipitable water vapor around 1.8 mm.
During these observations, the 183 GHz water line was monitored with water vapor radiometers (WVR), used to reduce atmospheric phase fluctuations. The quasars 
J1924$-$2914, J1820$-$2528, and J1831$-$2714 were used as the amplitude, atmosphere, bandpass, pointing, gain fluctuations, and WVR calibrators.

The continuum image presented in Figure \ref{fig:f1} was obtained by averaging line-free spectral channels of nine spectral windows. The total bandwidth for the continuum is about 3.75 GHz. 
The spectral windows were selected to observe different molecular lines as for example CH$_3$OH, HC$_3$N, CO, and SiO. In this paper, we concentrate in the emission of the SiO(5$-$4) and CO(2$-$1) 
spectral lines at rest frequencies of  217.105 and 230.538 GHz, respectively. An in-depth study for the rest of the lines will be presented in a future paper. The spectral windows where we detected the SiO and CO 
have a native channel spacing of 488.3 kHz or $\sim$0.61 km s$^{-1}$. However, given the broad velocity range of the lines, we smoothed the spacing channel to 2.44 km s$^{-1}$.  

The data were calibrated, imaged, and analyzed using CASA \citep{mac2007}, the Common Astronomy Software Applications package, Version 5.6. 
We also used some routines in Python to image the data \citep{ast2013}. We imaged the {\it uv}-data using the task TCLEAN.    
We set the {\tt\string Robust} parameter of TCLEAN equals to 0.5 for the continuum and line emissions.  We obtained an image {\it rms}-noise for the 1.3 mm continuum of 0.2 mJy Beam$^{-1}$ 
at an angular resolution of 0.12$''$ $\times$ 0.09$''$ with a PA of $-$75$^\circ$. The ALMA theoretical  {\it rms}-noise for this configuration, integration time, bandwidth, 
and frequency is approximately 0.02 mJy beam$^{-1}$, which is an order of magnitud smaller.  The strong continuum emission from the UCH$_{\rm II}$ region (with a density flux of about 10 Jy 
and a dynamic range of 200) did not allow us to reach the theoretical noise levels. An alternative possibility is the lack of short spacing information on the continuum structure.

The line image  {\it rms}-noise is 0.75 mJy beam$^{-1}$ km s$^{-1}$ at an angular resolution of 0.13$''$ $\times$ 0.09$''$ with a PA of $-$73$^\circ$. 
The ALMA theoretical rms noise for this configuration, integration time, bandwidth (channel spacing), and frequency is about 0.7 mJy beam$^{-1}$, which is also very close to the value we 
obtain in the line images. It is interesting to note that even when the line sensitivity is very close to that obtained theoretically, for the continuum emission it is not the case, again suggesting 
that the bright emission from the UCH$_{\rm II}$ region  is responsible for increased the {\it rms}-noise. 
 
 Phase self-calibration was done using the continuum emission as a model and then we applied the solutions to the line emission. We obtained about a factor two of improvement in the {\it rms}-noise. 

 \section{Results and Discussion} \label{sec:res}
 
 In Figure \ref{fig:f1} we present the CO(2$-$1) moment one map (intensity weighted velocity image), traditionally used to obtain the ``velocity field", overlaid with the 1.3 mm 
continuum emission,  and the 34 expanding filaments reported here that are related with this outflow. 
This map may help the reader to see clearly the expanding filaments emerging from the center of the flow.
This thermal emission was especially selected from outside the velocity 
range from $-$20 to $+$40 km s$^{-1}$, where the emission arises from the systemic molecular cloud and probably from some other clouds along the line-of-sight.  At these velocities there are 
also strong absorption features associated with the UCH$_{\rm II}$ at the center of the flow. We here avoid the confusing emission close to the embedding cloud systemic velocity, 
rather focussing on the filaments at higher velocities. The CO(2$-$1) thermal emission is found from $-$170 up to $+$130 km s$^{-1}$. These velocities are broader as compared to those radial velocities reported in recent SMA studies \citep{hun2008,su2012,zap2019}, probably because the better ALMA sensitivity allows us to detect fainter emission at higher velocities. We find 17 blueshifted, and 17 redshifted expanding filaments emerging from G5.89$-$0.39. 
Each filament marks the position of one sequence of molecular condensations mapped at different spectral channel velocities as already done in this, and other massive star forming regions \citep{zap2009,zap2013,zap2019}. 
These filaments coincide very well with some filamentary structures already traced by the moment one map and the six explosive filaments 
reported in \citet{zap2019}. The most prominent filaments in this Figure, the three blueshifted to the south/north of the explosive event and the two redshifted to the
north, are the ones already reported by those SMA observations \citep{hun2008,su2012,zap2019}. The rest of the filaments are new detections and give a more complete view of the explosiveness 
of the flow in G5.89$-$0.39.   

\begin{figure}
\centering
\includegraphics[scale=0.9, angle=0]{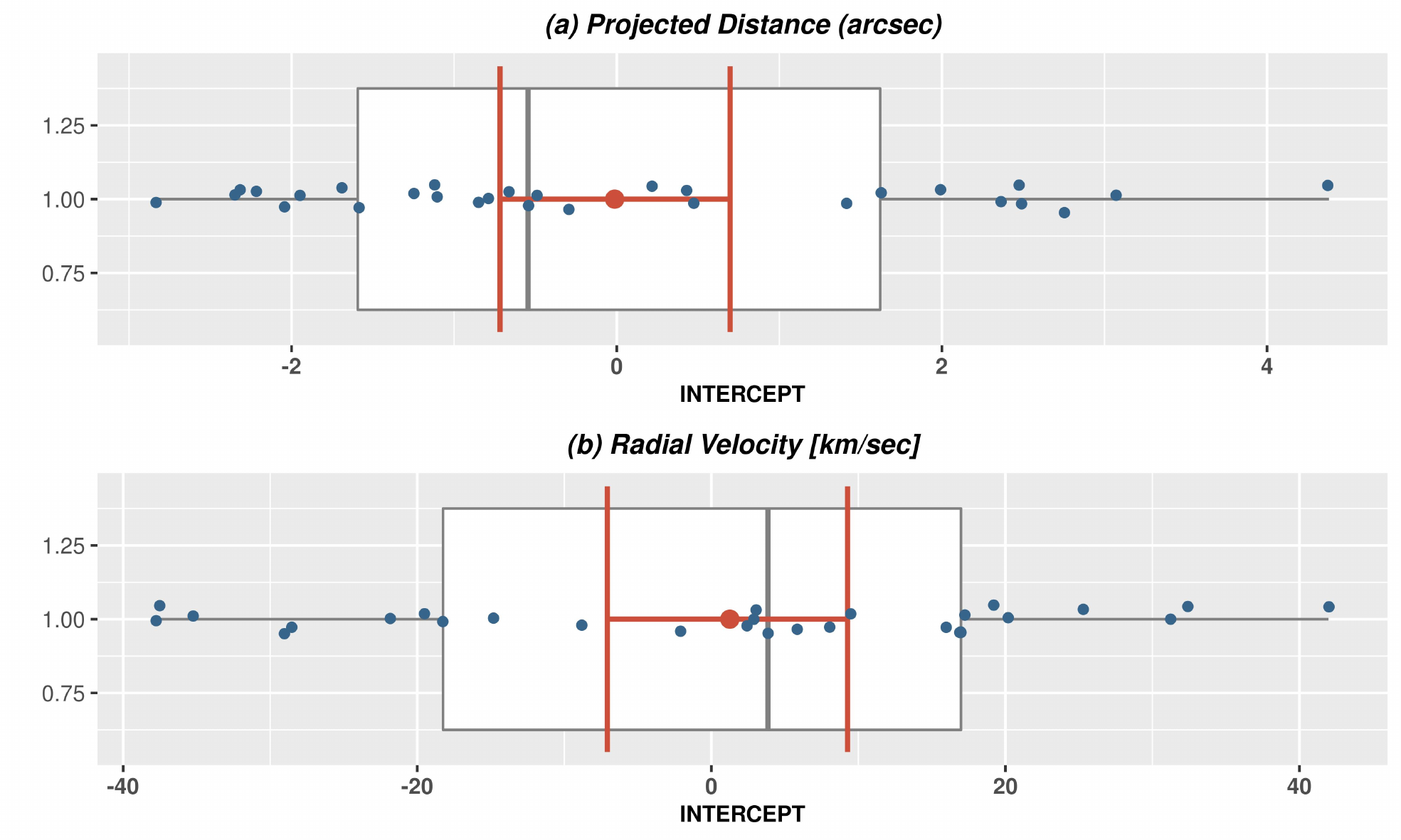}
\caption{ \scriptsize 95$\%$ intercept estimates confidence intervals (red lines) for the projected distance (a) with a statistical error of 0.7421$''$, and radial velocity (b) with statistical error of 8.5129 km s$^{-1}$. 
 The blue points represent  the jittered intercept estimated points. Relative origin (0,0) belongs to each interval with p-value estimated of 0.9725 and 0.7643.
 The grey line and the red dot represent the median and mean values, respectively.} 
\label{fig:f3}
\end{figure}   
 
\begin{figure}
\centering
\includegraphics[scale=0.6, angle=0]{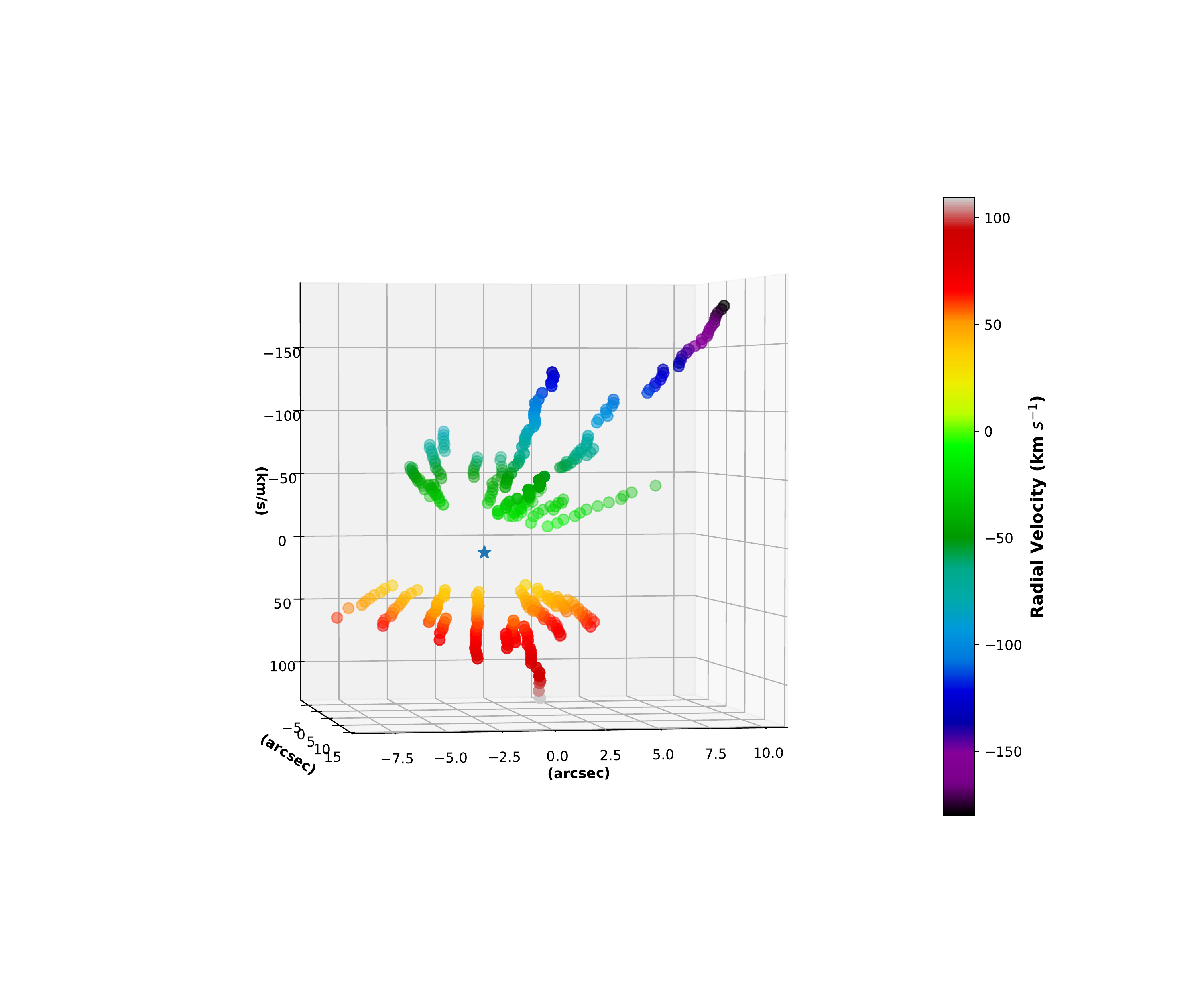}
\caption{ \scriptsize  Three-dimensional animation of the explosive event in G5.89$-$0.39. The radial blueshifted and redshifted velocities are shown from blue to red colors. 
The LSR radial velocity scale-bar (in km s$^{-1}$) is shown at the right. The (0$''$,0$''$,0 km s$^{-1}$) position is the origin. The star marks the position of the explosive outflow origin.
The animation starts with a view from up to down and then left to right. The duration of the animations is about 10 seconds. } 
\label{fig:f4}
\end{figure}

In Figure \ref{fig:f1}, it is also clear that the filaments get bluer or redder with increasing projected distance from their origin, suggesting that they follow a Hubble-velocity law, that is, the velocities 
increase linearly with on-the-sky distance from the center. This phenomenon is confirmed in the position-velocity diagram obtained in Figure \ref{fig:f2} or the 3D view 
presented in Figure \ref{fig:f4}. This property is one of the main features that characterize the explosive outflows \citep{zap2017}.  In Figure \ref{fig:f1}, 
the 1.3 mm continuum contours delineate a shell-like structure that is associated with the expanding and ionized UCH$_{\rm II}$ region reported in G5.89$-$0.39 \citep{woo1989,aco1998,hun2008}. 
All filaments converge to the center of the UCH$_{\rm II}$, the place where they originated. This position is located at  $\alpha_{J2000.0}$ = \dechms{18}{00}{30}{42} $\pm$ 0.005$^s$ 
and $\delta_{J2000.0}$ = \decdms{24}{04}{01}{5} $\pm$ 0.75$''$. The position given here is similar to the one obtained in \citet{zap2019} for the convergence of the six 
molecular filaments, but now with a better error.  In addition, we also include in this Figure the positions of the two objects located in the vicinities of the ionized shell, the Feldt \citep{fel2003}  and Puga \citep{pug2006} 
objects reported at infrared wavelengths. In this image, we mark the location of the dusty and compact objects reported by the SMA observations \citep{hun2008}.  
Massive young stars found in the periphery of the UCH$_{\rm II}$ region are likely related to the energetic explosion, but an in-depth proper motion 
study at optical or maybe infrared wavelengths is necessary to search for runaway massive stars in this region, as in the case of Orion BN-KL \citep{rod2020}. 
A search for an optical counterpart of the Feldt's star in the GAIA catalogue failed \citep{zap2019}, likely because of the high extinction at low galactic latitudes \citep{fos2012}. 

In Figure \ref{fig:f2},  we present the on-the-sky distance vs. the radial velocity plot of the 34 expanding filaments reported in this letter, and presented in the moment one map (Figure \ref{fig:f1}). 
As mentioned earlier, the filaments follow nearly straight lines and seem to converge to a radial velocity range between $-$7.2 to $+$9.7 km s$^{-1}$, the systemic velocity of the molecular cloud in 
G5.89$-$0.39 is around 9 km s$^{-1}$, see Figure 7 from \citet{hun2008}.  In all filaments the radial velocities increase linearly with the projected distance to the center of the UCH$_{\rm II}$ region, that is, 
follow a clear Hubble velocity-law. This pattern was already revealed by the SMA observations for the six filaments \citep{zap2019}, but it was not completely demonstrated that 
they belonged indeed to an explosive outflow.  The lack of deceleration in the filaments is an indication of the impulsive nature of the expansion.  
This implies that the density of the ejecta must be substantially larger than the medium through which they move.
In a Hubble flow crated by an explosion about 1,000 years ago, slower ejecta has moved a smaller distance than faster ejecta which has moved farther.

In Figure \ref{fig:f3}, we present a statistical analysis on the origin of the explosive outflow in G5.89. In this Figure is shown the jittered intercept estimated 
points and mean interval confidence for intercepts. Removing three outlier trajectories (or filaments), the mean of the projected distance intercepts is $\bar{x}=-0.01''$ and a 95\% confidence interval for 
the mean of intercepts is IC$_\mu$=($-$0.75$''$,0.73$''$). For the radial velocity mean of intercepts is $\bar{x}=$+$1.26$ km s$^{-1}$ 
and a 95\% confidence interval for the mean of intercepts is IC$_\mu$=($-$7.25 km s$^{-1}$, 9.77 km s$^{-1}$). 
With p-value of 0.9725 for the projected distance and p-value of 0.7643 for the radial velocity the hypothesis of $\mu=0$  cannot be rejected for both cases. 
These results show that statistically all trajectories have a common coordinated system origin. We included this origin in Figure \ref{fig:f1}.

\begin{figure}
\centering
\includegraphics[scale=0.55, angle=0]{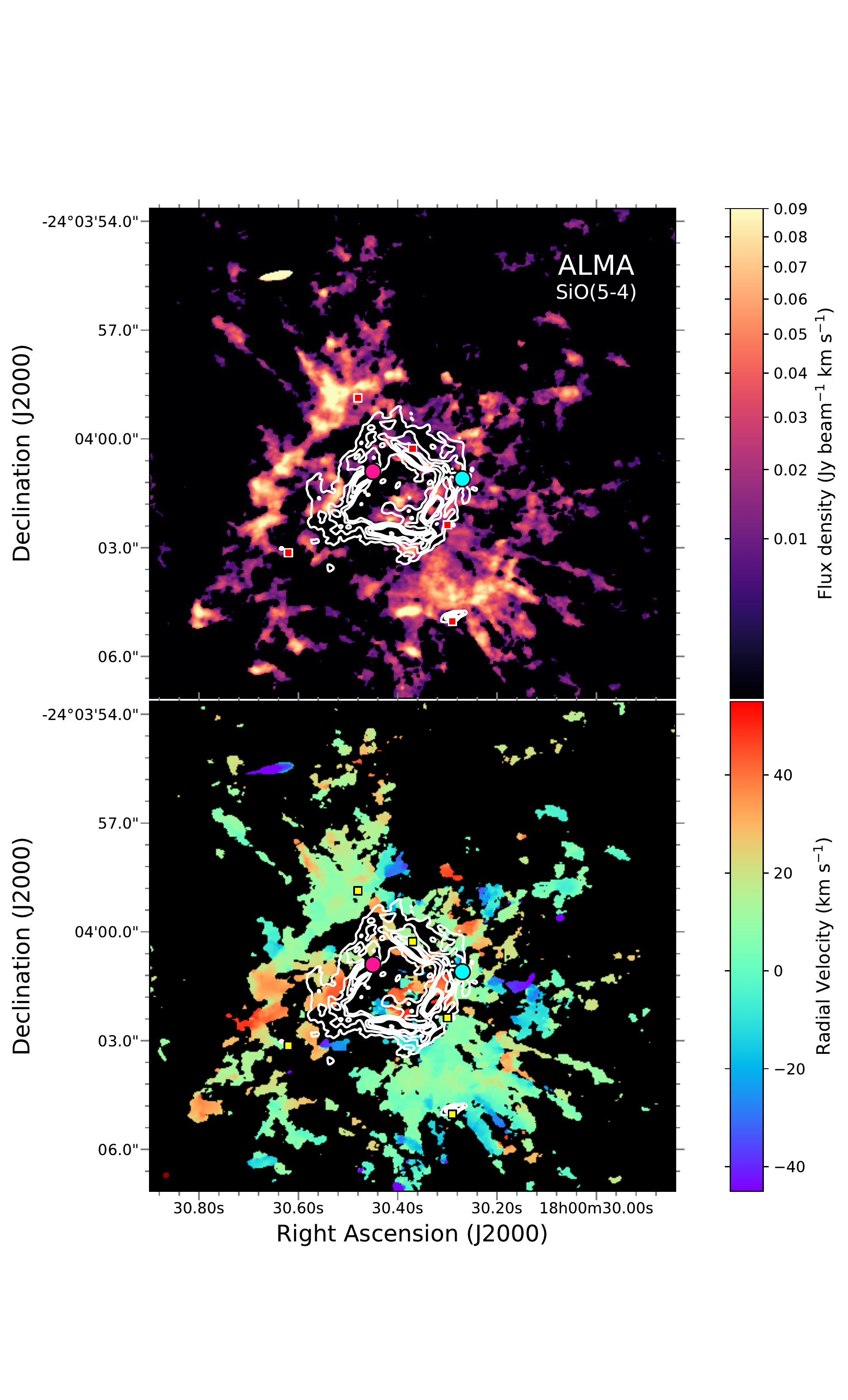}
\caption{ \scriptsize ALMA SiO(5-4) moment zero (upper panel) and one (lower panel) overlaid with the 1.3 mm continuum emission 
in white contours that are tracing the UCH$_{\rm II}$ region. The contours range from 10\% to 70\% of the peak emission, in steps of 10\%. 
The peak of the millimeter continuum emission is 34 mJy Beam$^{-1}$. 
To compute these maps, we integrated in radial velocities from $-$50 to  $+$50 km s$^{-1}$.
The half-power contour of the synthesized beam of the line image is shown in the bottom-left corner of the lower panel. The LSR radial velocity scale-bar (in km s$^{-1}$)
 is shown at the right of the lower panel. 
The location of the sources named Feldt's star \citep{fel2003} (pink circle) and Puga's star \citep{pug2006} (cyan circle) are shown  
at the center of the explosive outflow. The magenta squares mark the location of the objects reported with the SMA \citep{hun2008}.
All SMA objects have an ALMA counterpart, but this discussion is outside of the scope of the paper.
\label{fig:f5}}
\end{figure} 

The fact that the filaments are not straight lines reveals internal structure, probably because lateral or across-the-line-of-sight motions. 
From these discrepancies, we can estimate the linear dispersion across the filaments, assuming a displacement of about 1$''$ away from a linear trajectory over 1000 years,
 implies a traverse velocity of about 15 km s$^{-1}$. Alternatively, these displacements may result from deflection of the ejecta by dense clumps. 
 From the position-velocity diagram, taking maximum radial velocity of about 160 km s$^{-1}$ and the maximum projected distance of 12$''$, we obtained 
a kinematic age of 1000 yrs, a similar value for the fossil outflow reported by \citet{kla2006} and the estimated time to reach the present position of the Feldt's star.

A three-dimension (3D) view of the explosive outflow is presented in Figure \ref{fig:f4}. This image has three axis, the projected distance in RA and DEC (in arcseconds), 
and the radial velocity (in km s$^{-1}$) of the filaments. Here, the central position is at (0$''$,0$''$,$+$0 km s$^{-1}$), where in RA it is at \dechms{18}{00}{30}{42}  and DEC it is at \decdms{24}{04}{01}{5}.    
In the Figure it is clearly seen how the colors/velocities change getting bluer or redder  far from the origin indicating a Hubble-law velocity of each filament as already noted. It is important to mention 
that inside of the UCH$_{\rm II}$ region there is no emission from the filaments, they appear to start a bit farther from this position, see Figure \ref{fig:f2}. 

In Figure \ref{fig:f5} the moments of order zero (upper panel) and one (lower panel) emission of the SiO(5$-$4) spectral line from G5.89$-$0.39 are presented. To make this Figure, we integrated from
$-$50 to $+$50 km s$^{-1}$ in radial velocities. This velocity range includes the radial velocities close to the systemic cloud velocities ($+$9 km s$^{-1}$), but as this molecule
emission is found mainly in strong shocks \citep{sch1997}, ALMA could sample very well all these velocities, with no ambient cloud contamination. 
We resolved the SiO thermal emission, and reveal many filaments in almost every orientation ({\it i. e.}  with a nearly isotropic orientation), 
and again pointing back to the center of the UCH$_{\rm II}$ region. Overall the SiO maps confirm the explosion in this region. 
Only a few SiO filaments have a counterpart in the CO emission. Most of the CO filaments are found farther out from the SiO ones, and with broader velocity ranges.
The northwest bipolarity reported for this SiO outflow \citep{aco1998,soll2004} is totally lost in these new ALMA images, in which the outflow is more resemblant to an explosion.


Finally, one can make a very crude estimation of the rate of the 
explosive outflows occurring in our Galaxy using the case of Orion BN/KL , DR21, and  G5.89$-$0.39. If we assume that this event occurs approximately three times
every 10,000 years (the kinematic age of DR21) in a radius of 2.99 kpc, we obtain a rate of about one explosion every 130 years, close to the rate of supernovae \citep{tam1994}. 
The similarity between the event rates of explosive protostellar outflows and supernovae suggests that stellar dynamical interactions may play important roles in the formation
of massive stars. Models of high-mass star formation should be revised to include dynamical interactions.    

\section{Conclusions}

 In summary, the sensitive and high angular ALMA observations allow us to find that the outflow in G5.89$-$0.39 is indeed an explosion that emerged from the center of 
an ionized UCH$_{\rm II}$ region, and where young massive stars placed in its periphery could have powered the flow, but they moved from the center. 
G5.89$-$0.39 is thus the third explosive outflow in the Galaxy until now.  
Dedicated searches for explosive outflows in close-by massive star forming regions (for instance, if they take place when a protostellar merger is produced or by the formation of a 
capture of a companion into a close-binary orbit) could show this phenomenon to be more common than previously thought. 
       
\facilities{ALMA}
\software{CASA \citep{mac2007}, KARMA \citep{goo1996}}

\acknowledgments
We would like to thank the anonymous referee for the thoughtful suggestions that helped to improve  our manuscript.
This paper makes use of the following ALMA data: ADS/JAO.ALMA\#2018.1.00513.S. ALMA is a partnership of ESO (representing its member states), NSF (USA) and NINS (Japan), 
 together with NRC (Canada), MOST and ASIAA (Taiwan), and KASI (Republic of Korea), in cooperation with the Republic of Chile. The Joint ALMA Observatory is operated by ESO, AUI/NRAO and NAOJ.
 The National Radio Astronomy Observatory is a facility of the National Science Foundation operated under cooperative agreement by Associated Universities, Inc.
 L.F.R., is grateful to CONACyT, México, and DGAPA, UNAM for the financial support. L.A.Z. acknowledges financial support from CONACyT-280775 and UNAM-PAPIIT IN110618 grants, México. 
 A.P. acknowledges financial support from CONACyT and UNAM-PAPIIT IN113119 grant, México. P.T.P.H. acknowledges financial support from  MOST 108-2112-M-001-016-MY1.
 P.S. was partially supported by a Grant-in-Aid for Scientific Research (KAKENHI Number 18H01259) of the Japan Society for the Promotion of Science (JSPS).
 P.R. Rivera-Ortiz acknowledge funding from the European Research Council (ERC) under the European Union’s Horizon 2020 research and innovation programme, 
 for the Project “The Dawn of Organic Chemistry” (DOC), grant agreement No 741002.

\bibliographystyle{aasjournal}

\begin{thebibliography}{}
\bibitem[Acord et al.(1998)]{aco1998}  Acord, J.~M., Churchwell, E., \& Wood, D.~O.~S.\ 1998,ApJ, 495, L107
\bibitem[Astropy Collaboration et al.(2013)]{ast2013} Astropy Collaboration, Robitaille, T.~P., Tollerud, E.~J., et al.\ 2013, A\&A, 558, A33.
\bibitem[Bally et al.(2005)]{ball2005}  Bally, J. \& Zinnecker, H.\ 2005, AJ, 129, 2281
\bibitem[Bally et al.(2017)]{ball2017}  Bally, J., Ginsburg, A., Arce, H., et al.\ 2017, ApJ, 837, 60
\bibitem[Bally et al.(2020)]{ball2020} Bally, J., Ginsburg, A., Forbrich, J., et al.\ 2020, ApJ, 889, 178
\bibitem[Foster et al.(2012)]{fos2012} Foster, J.~B., Stead, J.~J., Benjamin, R.~A., et al.\ 2012, ApJ, 751, 157
\bibitem[Feldt et al.(2003)]{fel2003}    Feldt, M., Puga, E., Lenzen, R., et al.\ 2003, ApJL, 599, L91
\bibitem[Gooch(1996)]{goo1996} Gooch, R.\ 1996, Astronomical Data Analysis Software and Systems V, 101, 80
\bibitem[Hampton et al.(2016)]{ham2016} Hampton, E.~J., Rowell, G., Hofmann, W., et al.\ 2016, Journal of High Energy Astrophysics, 11, 1
\bibitem[Harvey et al.(1998)]{har1988}    Harvey, P.~M., \& Forveille, T.\ 1988, A\&A, 197, L19
\bibitem[Hunter et al.(2008)]{hun2008} Hunter, T.~R., Brogan, C.~L., Indebetouw, R., \& Cyganowski, C.~J.\ 2008, ApJ, 680, 1271
\bibitem[Klaassen et al.(2005)]{kla2006} Klaassen, P.~D., Plume, R., Ouyed, R., et al.\ 2006, ApJ, 648, 1079
\bibitem[McMullin et al.(2007)]{mac2007} McMullin, J.~P., Waters, B., Schiebel, D., Young, W., \& Golap, K.\ 2007, Astronomical Data Analysis Software and Systems XVI, 376, 127
\bibitem[Puga et al.(2006)]{pug2006}  Puga, E., Feldt, M., Alvarez, C., et al.\ 2006, ApJL, 641, 373
\bibitem[Rivera-Ortiz et al.(2019)]{riv2019} Rivera-Ortiz, P.~R., Rodr{\'\i}guez-Gonz{\'a}lez, A., Hern{\'a}ndez-Mart{\'\i}nez, L., et al.\ 2019, ApJ, 885, 104
\bibitem[Rodriguez et al.(2020)]{rod2020}  Rodr{\'\i}guez, L.~F., Dzib, S.~A., Zapata, L., et al.\ 2020, ApJ, 892, 82
\bibitem[Sato et al.(2014)]{sat2014}   Sato, M., Wu, Y.~W., Immer, K., et al.\ 2014, ApJ, 793, 72
\bibitem[Schilke et al.(1997)]{sch1997} Schilke, P., Walmsley, C.~M., Pineau des Forets, G., et al.\ 1997, A\&A, 321, 293
\bibitem[Sollins et al.(2004)]{soll2004} Sollins, P.~K., Hunter, T.~R., Battat, J., et al.\ 2004, ApJL, 616, L35
\bibitem[Su et al.(2009)]{su2009}   Su, Y.-N., Liu, S.-Y., Wang, K.-S., et al.\ 2009, ApJL, 704, L5
\bibitem[Su et al.(2012)]{su2012}   Su, Y.-N., Liu, S.-Y., Chen, H.-R., et al.\ 2012, ApJL, 744, L26
\bibitem[Tammann et al.(1994)]{tam1994} Tammann, G.~A., Loeffler, W., \& Schroeder, A.\ 1994, ApJS, 92, 487
\bibitem[Wood et al.(1989)]{woo1989} Wood, D.~O.~S., \& Churchwell, E.\ 1989, ApJS, 69, 831
\bibitem[Zapata et al.(2009)]{zap2009}  Zapata, L.~A., Schmid-Burgk, J., Ho, P.~T.~P., et al.\ 2009, ApJ, 704, L45
\bibitem[Zapata et al. (2013)]{zap2013} Zapata, L.~A., Schmid-Burgk, J., P{\'e}rez-Goytia, N., et al.\ 2013, ApJ, 765, L29
\bibitem[Zapata et al.(2017)]{zap2017}  Zapata, L.~A., Schmid-Burgk, J., Rodr{\'\i}guez, L.~F., et al.\ 2017, ApJ, 836, 133
\bibitem[Zapata et al.(2019)]{zap2019}  Zapata, L.~A., Ho, P.~T.~P., Guzm{\'a}n Ccolque, E., et al.\ 2019, MNRAS, 486, L15
\end{thebibliography}



\end{document}